\documentclass[fleqn,twoside]{article}
\usepackage{espcrc2}

\usepackage{graphicx}

\newcommand{\AmS}{{\protect\the\textfont2
  A\kern-.1667em\lower.5ex\hbox{M}\kern-.125emS}}
\newcommand{\tjusan}{\theta_{13}}

\title{The KASKA project -- a Japanese medium-baseline reactor-neutrino
       oscillation experiment to measure the mixing angle $\tjusan$ --}

\author{Masahiro Kuze\address{Department of Physics, Tokyo Institute of
        Technology,
        Tokyo, 152-8551 Japan}
        (for the KASKA Collaboration\thanks{The KASKA group institutes at the
         time of the workshop were:
KEK,
Kobe Univ.,
Niigata Univ.,
Rikkyo Univ.,
Tohoku Univ.,
Tokyo Met. Univ. and
Tokyo Tech.
})
               }       
\begin{document}

\begin{abstract}
A new reactor-neutrino oscillation experiment, KASKA, is proposed
to measure the unknown neutrino-mixing angle $\tjusan$ using the world's
most powerful Kashiwazaki-Kariwa nuclear power station.
It will measure a very small deficit of reactor-neutrino flux
using three identical detectors, two placed just close to
the sources and one at a distance of about 1.8km.
Its conceptual design and physics reach are discussed.
\vspace{1pc}
\end{abstract}

\maketitle

\section{Introduction}
The investigation of the mixing phenomena of the leptonic sector will
be a major theme of particle physics in next few decades, as
has been that of the quark mixing in the past.
While two neutrino-mixing angles, $\theta_{12}$ and $\theta_{23}$,
have been measured to be fairly large~\cite{SK,SNO,K2K,KamLAND},
only upper limits have been obtained for the remaining angle $\tjusan$.
The best limit is $\sin^2 2\tjusan<0.2$ for $\Delta m^2=2.0\times
10^{-3}~\rm{eV}^2$ from CHOOZ~\cite{CHOOZ}, and this smallness in
contrast to the other angles is a mystery that challenges in building up
a theory to explain the origin of the mixing.  Not only the measurement of
the angle itself is of high scientific interest, but also knowing the range
of this parameter is a crucial input in planning for future
very-long-baseline oscillation experiments to detect the leptonic
CP-violating phase $\delta_{CP}$, using very intense beams of
neutrinos (and anti-neutrinos) from Superbeam facilities or
Neutrino Factories.

It has been shown~\cite{minakata}, also nicely in the plenary session
of this workshop~\cite{thierry}, that the detection (or setting new
upper limit) of $\tjusan$ by $\bar{\nu_e}$ disappearance experiments
with reactor neutrinos gives complementary information to that from the
planned $\nu_\mu \to \nu_e$ appearance experiments with accelerator
neutrino beams.  Therefore, having both kinds of experiments,
preferably concurrently, would be useful in the exploration of
the mixing phenomena in the lepton sector.

The KASKA experiment will try to achieve the mission using
reactor neutrinos from KAShiwazaki-KAriwa nuclear power station
in Niigata, Japan.  Its conceptual detector design, background and
systematic-uncertainty consideration and physics reach will be discussed
in the following.

\section{The experiment}
Kashiwazaki-Kariwa nuclear power station, owned and operated by
Tokyo Electric Power Company, is the biggest source of reactor
neutrinos in the world, with a power of 23.4~GWth.
The station has seven reactors, which are clustered into four and three.
The two clusters are separated by about 1.5~km in distance.
The purpose of the KASKA experiment is to detect a small
deficit of anti-neutrinos due to $\tjusan$ mixing:
\begin{equation}
P_{\bar \nu_e \rightarrow \bar \nu_e} =1-\sin^22\tjusan \sin^2\frac
{\Delta m^2_{13} L}{4E_{\nu}} + \boldmath{\mathcal {O}}(10^{-3}).
\label{eq1}
\end{equation}  
Note that the value of $\Delta m^2_{13}$ is very close to
$\Delta m^2_{23}$ measured by SK and K2K, as $\Delta m^2_{23}
>> \Delta m^2_{12}$.
With the neutrino energy considered (the detected neutrino
energy peaks at 4~MeV), the oscillation maximum occurs at
1-2~km, and other effects (last term in Eq.~\ref{eq1})
become negligible.
Therefore, the experiment is a pure $\tjusan$ measurement and
detects a small deficit due to $\tjusan$ and $\Delta m^2_{13}$,
while the KamLAND experiment measured a large deficit due to
$\theta_{12}$ and $\Delta m^2_{12}$ at a distance of about 180~km.
%
%
In order to minimize the systematic uncertainties related to the absolute
neutrino-flux estimation and absolute detection efficiency,
it is planned to have multiple detectors,
ones close to the source (near detectors) and the other close to the
oscillation maximum (far detector).  
KASKA will have two near detectors for the two reactor clusters,
each at a distance of approximately 400~m.  The location of the 
far detector will be at about 1.8~km from both clusters,
corresponding to the $\Delta m^2_{23}$ value measured at SK and K2K.

\begin{figure}[tb]
\includegraphics[width=20pc]{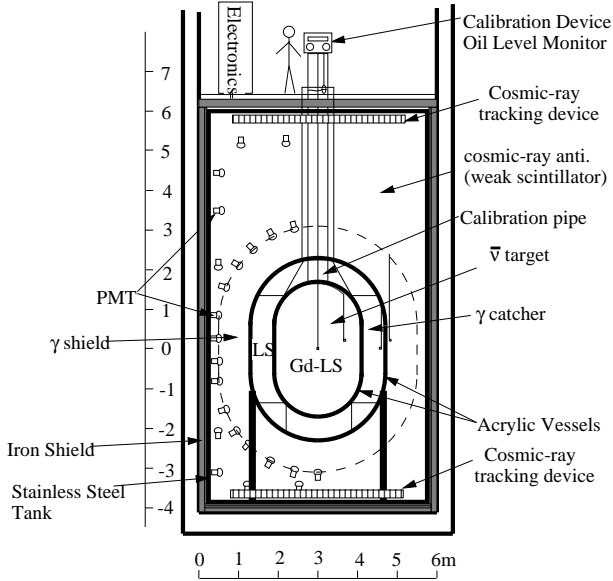}
\caption{Schematic view of the KASKA detector.
\label{fig:detector}}
\end{figure}

Figure~\ref{fig:detector} shows the schematic view of the planned
detector.  It consists of the central target region and a few
buffer and veto layers surrounding it.
The neutrino target is 8 tons of liquid scintillator (LS) contained
in an acrylic vessel.  The LS is loaded with 0.1\% of Gadolinium (Gd).
The detection is via inverse-$\beta$ reaction,
$\bar{\nu}_e + p \rightarrow e^+ + n$,
which produces a prompt signal due to the positron energy loss
and its annihilation with an electron.
As the recoil energy to the nucleon is very small, it gives
a measurement of the incoming neutrino energy.
Then the neutron is captured by Gd, which emits several
$\gamma$-rays whose total energy amounts to 8~MeV.
This gives a delayed signal, and taking a coincidence
of the two signals in time sequence (the typical capture time
is 30~$\mu s$)
greatly reduces the background.

The target region is surrounded by the $\gamma$-catcher region,
which is a 60-cm thick layer of LS without Gd loading,
but with the same light output as the target LS.  It detects
$\gamma$ rays that escaped from the target region, thus improving
the energy reconstruction of the prompt and delayed signals.

Out of the $\gamma$ catcher is the buffer region, a 90-cm layer
of non-scintillating paraffin oil.  This works as a shield against
background $\gamma$ and $\beta$ rays from radioactive elements
contained in the photomultiplier (PMT) glasses.
The PMTs are attached at the outer surface of this region.

The outermost layer is the cosmic-veto region, a layer of
LS with weak light output.  This layer will be viewed
by PMTs attached to the outermost part of the detector.
It detects the cosmic muons entering the detector, giving
a veto signal against background events associated with them.

In addition, there will be cosmic-ray tracking devices
at the top and bottom of the detector.  They will reconstruct
the passage of the cosmic rays and will be used to estimate
the background coming from muon spallation products.

In order to reduce the cosmic-induced background to a very
low level, the detectors will be located in underground
vertical shafts.  The near detectors will be placed at a depth
of about 70~m.  The shaft for the far detector will be about
200~m deep, due to the smaller signal event rate.
The expected cosmic-muon rates for the entire detector volume
are roughly 100~Hz for the near detectors and 10~Hz for the far detector.

Following sources of background are considered:
\begin{itemize}
\item{Accidental coincidence of $e^+$-like singles rate and neutron-like
singles rate.}
\item{Fast neutrons, mimicking a prompt signal (by recoiling a proton,
for example) and then captured, thus producing a correlated background.}
\item{Long-lived spallation products, such as $^9\rm{Li}$ or $^8\rm{He}$,
induced by cosmic ray.  Their decay involves $\beta$ and neutron emissions,
thus constituting a correlated background.}
\end{itemize}
With enough shaft depth and shielding, it is possible to keep the
background at the level of 1\% of the signal, with an uncertainty of
0.3\% or less.

\begin{figure}[tb]
\includegraphics[width=20pc]{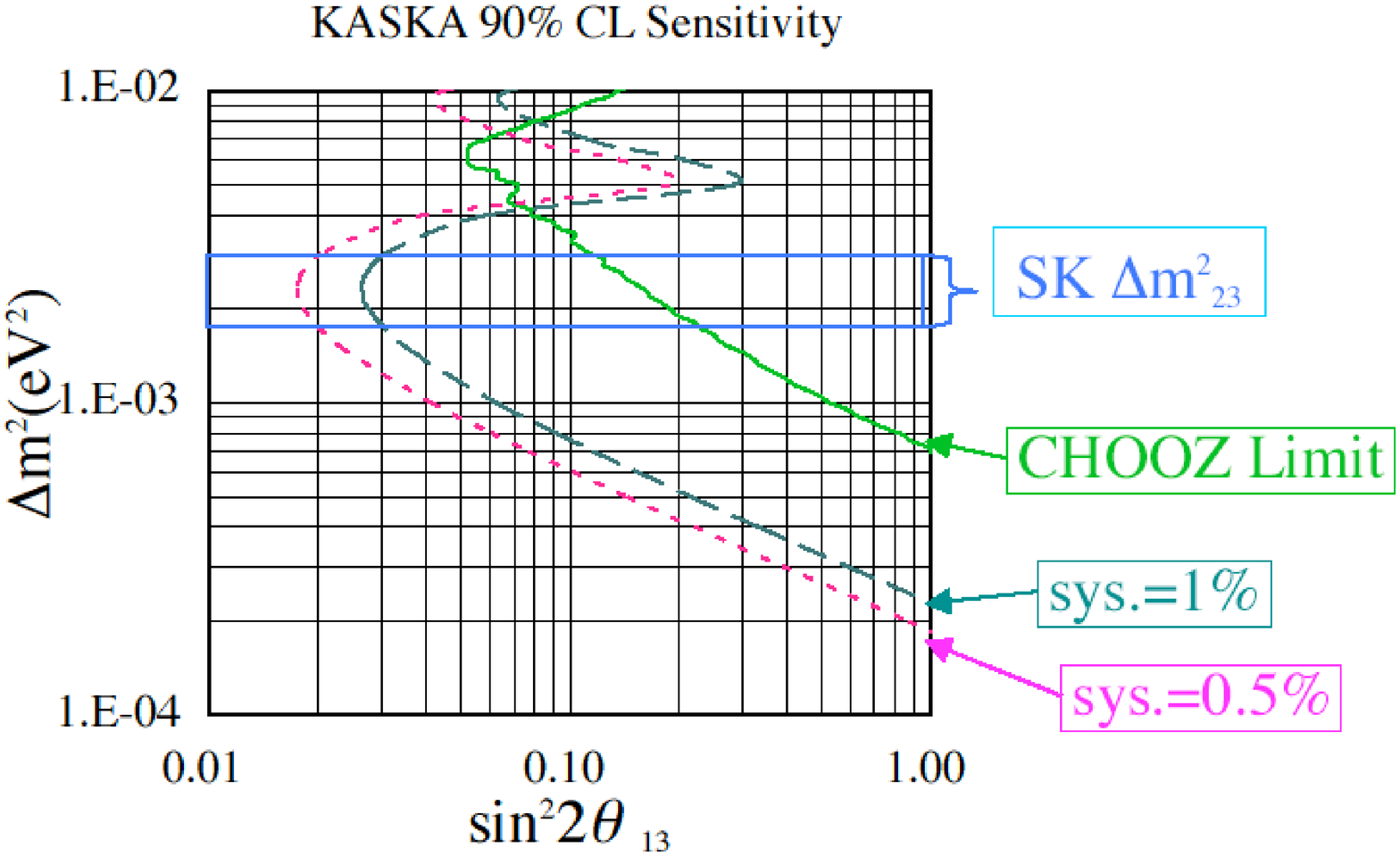}
\caption{ KASKA 90~\% C.L. sensitivity after three years of running.
The $\Delta m^2_{23}$ region allowed by Super-Kamiokande~\cite{SK}
at 90~\% C.L. is also indicated.
\label{fig:sensitivity}}
\end{figure}

With a running period of three years, the far detector will collect
about 30,000 neutrino events, i.e. with a statistical error of 0.6\%.
During this period, the near detectors will collect 300,000 events
each, providing also inputs for reactor science using the very
precise neutrino $\beta$-spectrum.  The main source of systematic
uncertainty comes from the relative acceptance difference between
the near and far detectors, originating, for instance, from the
target volume.  The analysis of the systematic sources can be found
elsewhere~\cite{suekane}.  It is possible to control the systematic
uncertainty well below 1\%.
The uncertainty related to the absolute neutrino-flux estimation,
which was fairly large in the earlier-generation reactor experiments,
becomes negligible thanks to the multi-detector configuration.

Figure~\ref{fig:sensitivity} shows the expected sensitivity~\cite{suekane}
on $\sin^2 2\tjusan$ as a function of $\Delta m^2$, which improves
on CHOOZ limits by one order of magnitude.

\section{Status and prospects}
The collaboration currently consists of seven Japanese institutes,
and is growing.  Two-year R\&D funding was approved and has started
from FY 2004.  The R\&D includes 1) boring study of geologies at the
location of one of the near detectors, followed by in-situ background
measurements (cosmic rates and $\gamma$ backgrounds) at the bottom of the
70~m bore shaft, 2) building of a prototype detector (1.2-m diameter
sphere) to study the energy-reconstruction systematics, and 3)
development of readout electronics.  With more institutes joining,
further tests and developments are envisaged.  The negotiations
with the electric company and authorities have been progressing
successfully.  With an appropriate funding profile, the earliest
schedule could be to start detector and shaft constructions in 2006
and to start data-taking in 2008.

\end{document}